# Interferometric Identification of a Pre-Brown Dwarf

Philippe André[1]*, Derek Ward-Thompson[2], Jane Greaves[3]

[1]Laboratoire AIM, CEA/DSM-CNRS-Université Paris Diderot, IRFU/Service d'Astrophysique, C.E. Saclay, Orme des Merisiers, 91191 Gif-sur-Yvette Cedex, France.

[2]Jeremiah Horrocks Institute, University of Central Lancashire, PR1 2HE, UK.

[3]SUPA, Physics and Astronomy, University of St Andrews, North Haugh, St Andrews, Fife KY16 9SS, UK.

* To whom correspondence should be addressed. E-mail: philippe.andre@cea.fr

**Abstract**: It is not known whether brown dwarfs (stellar-like objects with masses less than the hydrogen-burning limit, 0.075 $M_\odot$) are formed in the same way as solar-type stars or by some other process. Here we report the clear-cut identification of a self-gravitating condensation of gas and dust with a mass in the brown-dwarf regime, made through millimeter interferometric observations. The level of thermal millimeter continuum emission detected from this object indicates a mass ~ 0.02-0.03 $M_\odot$, while the small radius < 460 AU and narrow spectral lines imply a dynamical mass of 0.015-0.02 $M_\odot$. The identification of such a pre-brown dwarf core supports models according to which brown dwarfs are formed in the same manner as hydrogen-burning stars.

Brown dwarfs, defined as stellar-like objects with masses less than the hydrogen-burning limit $M_{BD} = 0.075$ $M_\odot$ ([1]), were first discovered in 1995 ([2-3]). They are now known to be almost as numerous as hydrogen-burning stars ([4-6]), but their formation mechanism remains a matter for debate ([6-7]). Either brown dwarfs form as a by-product of the formation process of hydrogen-burning stars; or they form just like normal stars ([7]), from the collapse of self-gravitating condensations of gas and dust called prestellar cores ([8]). The models in the former category include: models of multiple star formation where the lowest mass member is ejected before accreting too much mass ([9-10]); models of circumstellar disk fragmentation ([11-12]); and models in which more massive prestellar cores are disrupted by nearby massive star formation into low-mass cores, which are then triggered into collapse ([13]). The type of models we test here are those in which brown dwarfs form like solar-mass stars, but are just less massive, such as in the gravo-turbulent fragmentation scenario ([14-15]). In this theoretical scenario, gas cores with a wide range of masses are formed from material compressed by shocks resulting from supersonic interstellar turbulence. At low masses, this includes both gravitationally bound and unbound cores. A fraction of the low-mass cores produced by turbulence are dense enough to be gravitationally unstable to collapse: their masses exceed the local Jeans or Bonnor-Ebert mass and are comparable to their virial mass. Indeed, turbulence can locally generate sufficient compression and background pressure ($P_{back}/k_B > P_{BD,10K}/k_B \approx 2.4 \times 10^7$ K cm$^{-3}$), that the corresponding critical Bonnor-Ebert mass, $M_{BE} = 0.12$ $M_\odot \times (T/10$ K$)^2 \times (P_{back}/10^7$ K cm$^{-3})^{-1/2}$ which has a typical value of $\sim 1$ $M_\odot$ in molecular clouds ([16]), becomes smaller than the maximum brown-dwarf mass $M_{BD} = 0.075$ $M_\odot$. This scenario predicts a large number of pre-brown dwarfs, which can be approximately described as compact Bonnor-Ebert isothermal spheres of mass $M_{BE} < M_{BD}$ and radius $R_{BE} < R_{BD} \approx 700$ AU ([17]), but no good example of such a pre-brown dwarf core has previously been confirmed observationally to be sufficiently compact to be self-gravitating.

Nearby, high-pressure cluster-forming regions such as L1688 in Ophiuchus [120-140 pc from the Sun ([18-19])] are the best places to search for pre-brown dwarfs. Following up on an early DCO$^+$ search ([20]), deep dust continuum observations with the SCUBA camera on the James Clerk Maxwell Telescope (JCMT) led to the detection of a candidate pre-brown dwarf in L1688 as an unresolved 850 μm point source with $S_{850\mu} = 39 \pm 5$ mJy/15''-beam ([21]). This candidate, called Oph B-11, is located in a region of relatively high visual extinction ($A_V \sim 30$ mag based on 2MASS) between the dense clumps Oph B1 and Oph B2 of L1688 [see ([22]) and Fig. S1].

We used the Institut de Radioastronomie Millimétrique (IRAM) Plateau de Bure interferometer (PdBI) in CD configuration to observe Oph B-11 at high spatial resolution at 3.2 mm, for a total of four interferometer tracks between March 2006 and April 2011. We obtained a robust detection of the source in both the continuum at 3.2 mm (Fig. 1) and the N$_2$H$^+$(1-0) line (Fig. 2 and Fig. S2). The 3.2 mm continuum PdBI source is located less than 2" away from the previously detected SCUBA position ([21]) (which has an uncertainty of +- 3''). It has a peak flux density $S_{3.2mm, peak} \approx 0.32 +- 0.05$ mJy/7.8''×2.8" beam ($> 6\sigma$ detection) and an integrated flux density $S_{3.2mm, tot} = 0.4 \pm 0.1$ mJy. The probability of chance coincidence with an extragalactic object is $< 0.05\%$ [see Supporting Online Material (SOM)]. The source is point-like at the PdBI resolution. Fitting a spherical Bonnor-Ebert core model ([23]) to the interferometric visibilities in the uv plane yields a best-fit value of 1.0"+-1.8" for the outer angular radius. This corresponds to a best-fit core radius $R_{core} \sim 140$ AU if we adopt a distance d = 140 pc for the Ophiuchus cloud (*19*).

The interferometric visibility curve sets an upper limit of 3.3" to the outer angular radius at the 90% confidence level (see Fig. 3). This constrains the outer radius of the Bonnor-Ebert core model to be $R_{core} < 460$ AU. The presence of an extended power-law envelope around the compact core is ruled out by the shape of the visibility curve (Fig. 3) and the low value of the 850 μm flux density measured in a 15" beam at JCMT (*21*).

We know that the Oph B-11 core is starless because it was not detected in deep mid-infrared observations of L1688 with *Spitzer* (*24*). It also remained undetected at both 70 μm and 100 μm in recent *Herschel* observations of L1688 made as part of the *Herschel* Gould Belt survey (*25*). Despite the high level of background emission from the L1688 cloud, the *Herschel* 70 μm data set a 3σ upper limit of 85 mJy to the flux density of Oph B-11 at 70 μm, which is significantly lower than the 70 μm flux density of ~ 200-300 mJy expected for a first hydrostatic protostellar core at the distance of Ophiuchus (*26-27*). Therefore, it is safe to conclude that there is no protostellar object at the center of Oph B-11. The available spectral energy distribution of Oph B-11 is consistent with emission from cold (T ≲ 10 K) dust (Fig. 4). Furthermore, the ambient radiation field in the L1688 cloud is relatively well constrained (*28*) and radiative transfer calculations (*29*) confirm that the core temperature should be as low as ~ 8.5-10 K given the high visual extinction $A_V \sim 30$ mag observed in the immediate vicinity of Oph B-11.

Assuming a typical dust opacity law for starless cores (*30-31*) (i.e., an opacity per unit gas+dust mass column density $\kappa_{850\mu} \sim 0.01$ cm$^2$ g$^{-1}$, and a dust emissivity index β = 2), the observed 850 μm (*21*) and 3.2 mm flux densities of Oph B-11 correspond to a mass of gas and dust $M_{obs} \sim 0.02 - 0.03$ M$_\odot$ at d = 140 pc for a dust temperature in the range of 8.5-10 K (see SOM for a discussion of the uncertainties). Adopting the upper limit of 460 AU to the core radius, the lower limits to the mean density and mean column density are estimated to be $<n_{H2}> \gtrsim 7.5 \times 10^6$ cm$^{-3}$ and $<N_{H2}> \gtrsim 6.9 \times 10^{22}$ cm$^{-2}$, respectively. Therefore, Oph B-11 has a minimum mean column density contrast of a factor ≥ 2.3 over the local background which has a typical column density ~ $3 \times 10^{22}$ H$_2$ cm$^{-2}$ (Fig. S1).

The linewidth measured for the N$_2$H$^+$(1-0) multiplet is extremely narrow (cf. Fig. 2), indicating that the total (thermal + nonthermal) one-dimensional velocity dispersion $\sigma_{tot}$ within the core is nearly thermal and comparable to the isothermal sound speed ($c_s \lesssim 0.2$ km/s for a gas temperature T ≲ 10 K). For a truncated $\rho \propto r^{-2}$ density distribution, the nominal virial mass or dynamical mass of Oph B-11 is estimated to be $M_{vir} = 3R_{core}\sigma_{tot}^2/G = 0.014-0.017$ M$_\odot$ for the best-fit core radius and T = 8.5-10 K. If the core radius is as large as the upper limit of 460 AU, then the virial mass reaches an upper limit of < 0.056 M$_\odot$. The virial mass ratio $\alpha_{vir} \equiv M_{vir}/M_{obs}$ of the Oph B-11 core thus has a nominal value of 0.5-0.9 and an upper limit of 2.8. On the theoretical side (*32*), $\alpha_{vir} \gg 2$ when self-gravity is unimportant, $\alpha_{vir} \lesssim 2$ for gravitationally-bound objects, and $\alpha_{vir} \lesssim 1$ for strongly self-gravitating objects in which gravitational energy dominates over kinetic energy. Both the derived value of $\alpha_{vir}$ and the high column density contrast strongly suggest that Oph B-11 is gravitationally bound. Because the present values of $M_{obs}$ and $M_{vir}$ are both smaller than the brown dwarf limit of 0.075 M$_\odot$ and the estimated final mass of the core is also substellar (see SOM), we conclude that Oph B-11 is a pre-brown dwarf.

Our interferometric observations thus demonstrate that pre-brown dwarfs do exist and therefore tend to support models such as the gravo-turbulent fragmentation picture (*15, 17*). However, they do not yet prove that pre-brown dwarfs are the main channel of brown dwarf formation. Although Oph B-11 itself is unlikely to have been ejected from a disk (see SOM), our results do not rule out the possibility that some brown dwarfs form by disk fragmentation and/or ejection (*9-12*).

**Acknowledgments:** This work is based on observations carried out with the IRAM PdBI. IRAM is supported by INSU/CNRS (France), MPG (Germany), and IGN (Spain). It was stimulated by discussions held in the context of the European Marie Curie Research Training Network `Constellation' (MRTN-CT2006-035890). We are grateful to the IRAM Director who awarded us 1 track of Director's Discretionary Time on the PdBI to confirm the detection of Oph B-11 in the 3.2 mm continuum. We thank F. Gueth and R. Neri for their help and expert advice concerning the interferometric data reduction for this project. We also thank M. Pound for useful comments on Oph B-11. The interferometric data used in this paper are archived at IRAM and available upon request from either IRAM or the corresponding author.

**Supporting Online Material (SOM):**

Supplementary Text
Figs. S1 to S2
References *(33 to 48)*

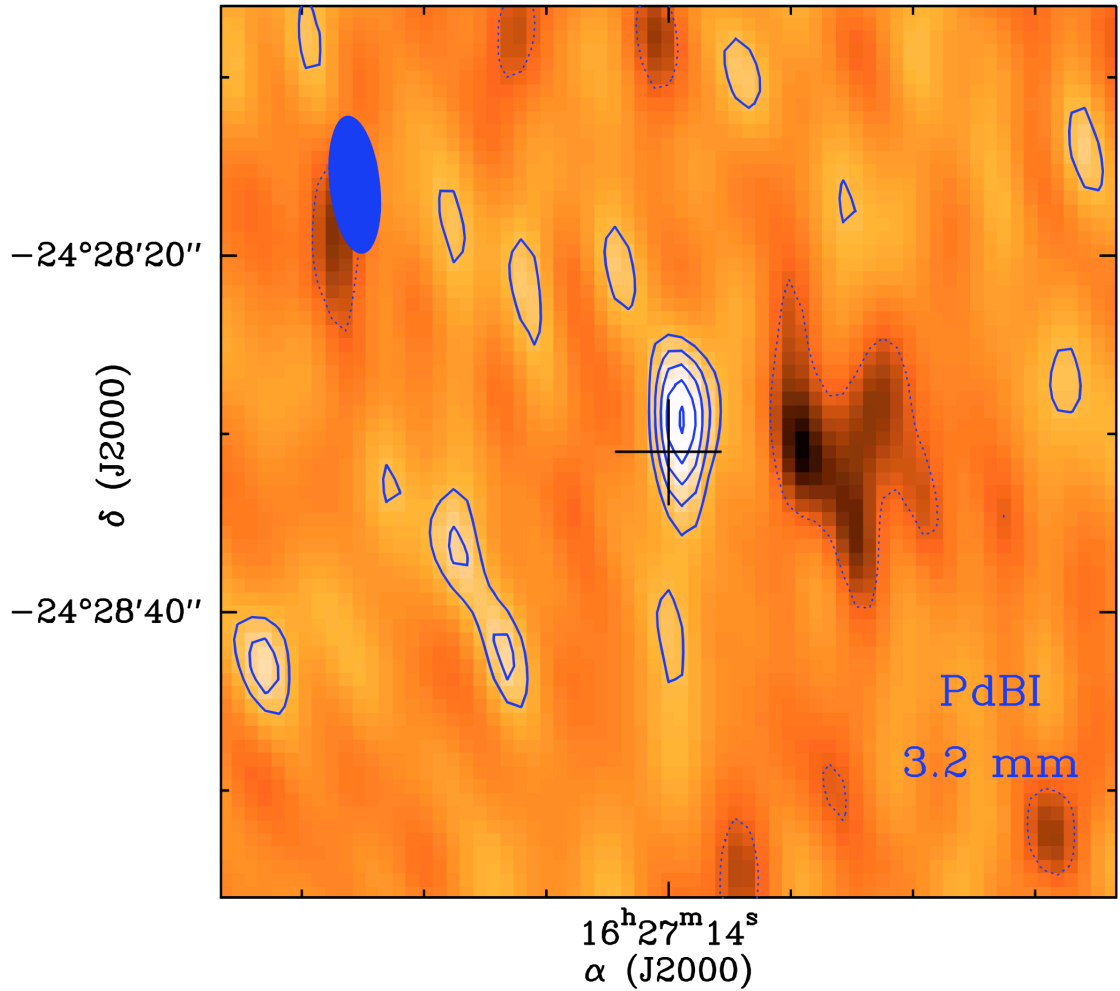

**Fig. 1.** Interferometric 3.2 mm continuum image of the candidate pre-brown dwarf Oph B-11. The angular resolution (or synthesized half-power beamwidth) achieved with the CD array of the IRAM PdBI is 7.8''×2.8'' as shown at the upper left. The rms noise is $\sigma \approx 0.05$ mJy/beam; the contours are $-2\sigma$ (dotted), $2\sigma$, $3\sigma$, $4\sigma$, $5\sigma$, and $6\sigma$. The SCUBA position (*21*) of Oph B-11 is marked by a cross whose size reflects the +-3'' pointing uncertainty of JCMT. The PdBI continuum peak is located at RA(J2000) = $16^h27^m13^s.96$ +- $0.02^s$, DEC(J2000) = $-24°28'29.3''$ +- 0.6'', or at $\Delta\alpha = -0.6''$+-0.2'', $\Delta\delta = +1.75''$+-0.6'' with respect to the SCUBA position. It has a peak flux density $S_{3.2mm, peak} \approx 0.32$ mJy/beam ($6.2\sigma$ detection).

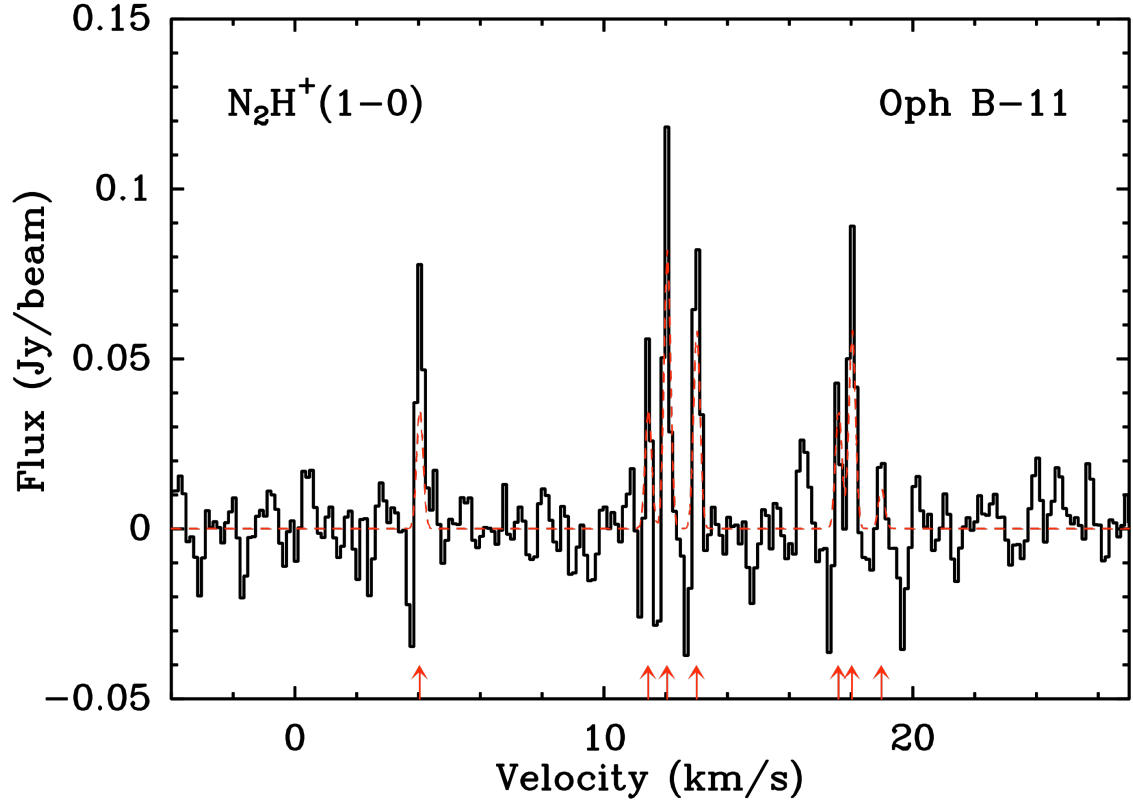

**Fig. 2.** Spectrum of Oph B-11 in the $N_2H^+(1-0)$ line multiplet. The data (black histogram) correspond to the beam-averaged $N_2H^+(1-0)$ spectrum observed with the D array of the IRAM PdBI at the position of the 3.2 mm continuum peak of Oph B-11 (see Fig. 1). The x-axis represents velocity relative to the local standard of rest (LSR) assuming a rest frequency of 93176.265 MHz for the isolated component of the line multiplet $N_2H^+(101-012)$. A Gaussian fit to the $N_2H^+(1-0)$ multiplet (superimposed as a dashed red curve) yields a full-width-at-half-maximum (FWHM) linewidth $\Delta v_{obs} = 0.21 \pm 0.03$ km/s, while the velocity resolution is $\Delta v_{res} = 0.125$ km/s. The expected positions of the seven components of the $N_2H^+(1-0)$ line multiplet are marked by red arrows. The fit implies a deconvolved FWHM linewidth $\Delta v_{int} = 0.17 \pm 0.04$ km/s, corresponding to a nonthermal velocity dispersion $\sigma_{NT} = 0.05 \pm 0.02$ km/s. The systemic velocity of the source is $V_{LSR} = 4.03 \pm 0.04$ km/s, similar to that of other objects in the dense clumps Oph B1 and Oph B2 of the Ophiuchus protocluster (*22*).

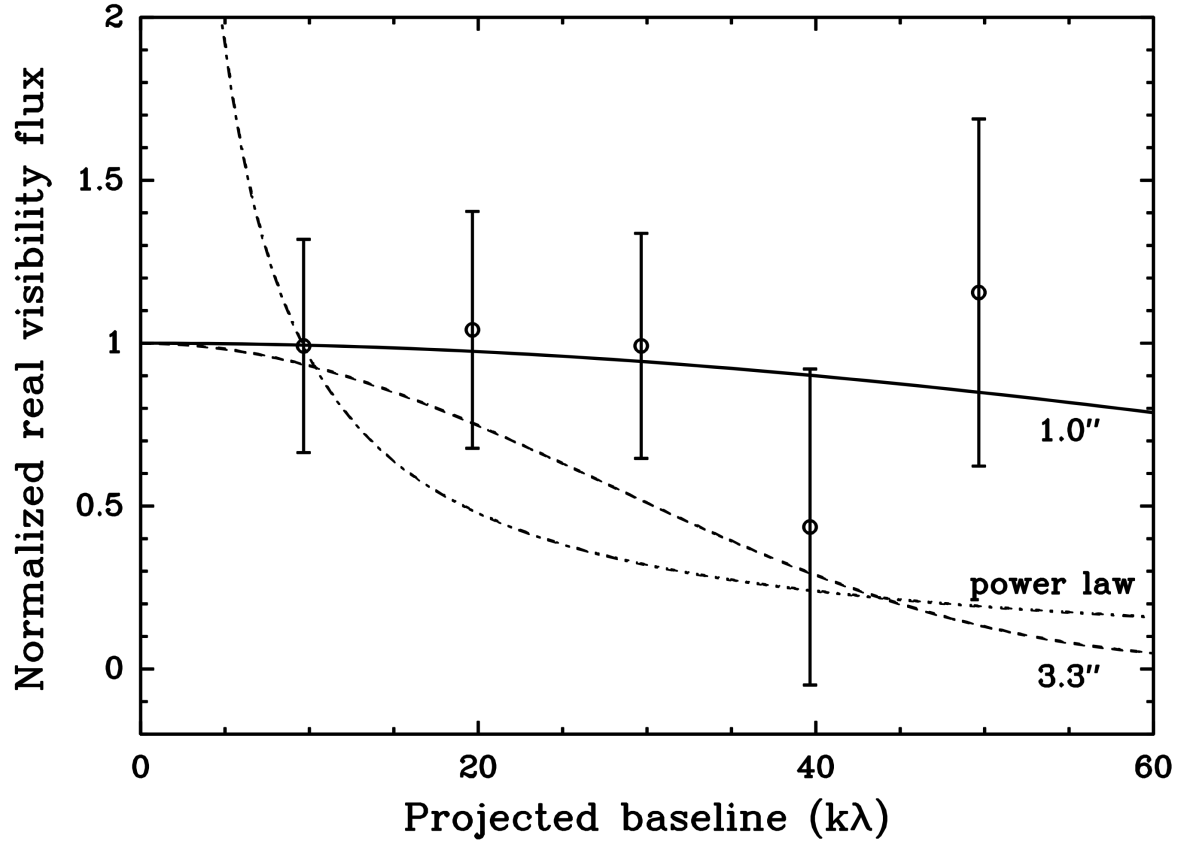

**Fig. 3.** Interferometric visibility curve of Oph B-11 compared to three model cores. The normalized 3.2 mm continuum visibility data correspond to the IRAM PdBI image shown in Fig. 1 and the error bars are +-1$\sigma$. The solid curve shows the best-fit model for a critical Bonnor-Ebert core, which has an outer angular radius of 1.0″. The dashed curve corresponds to a Bonnor-Ebert core model with an outer angular radius of 3.3″, which is the largest radius consistent with the uv-plane data at the 90% confidence level. This confidence level was estimated by comparing the reduced chi square of the 3.3″ model, which is 2.14 compared to 0.44 for the best-fit model, to the chi-square probability distribution for 3 degrees of freedom (5 normalized visibility points, i.e., 4 independent data points, and 1 free parameter: the model outer radius). The dash-dotted curve corresponds to an extended power-law ($\rho \propto r^{-2}$) model, normalized to the first visibility point. It is ruled out by the shape of the visibility curve at the 95% confidence level.

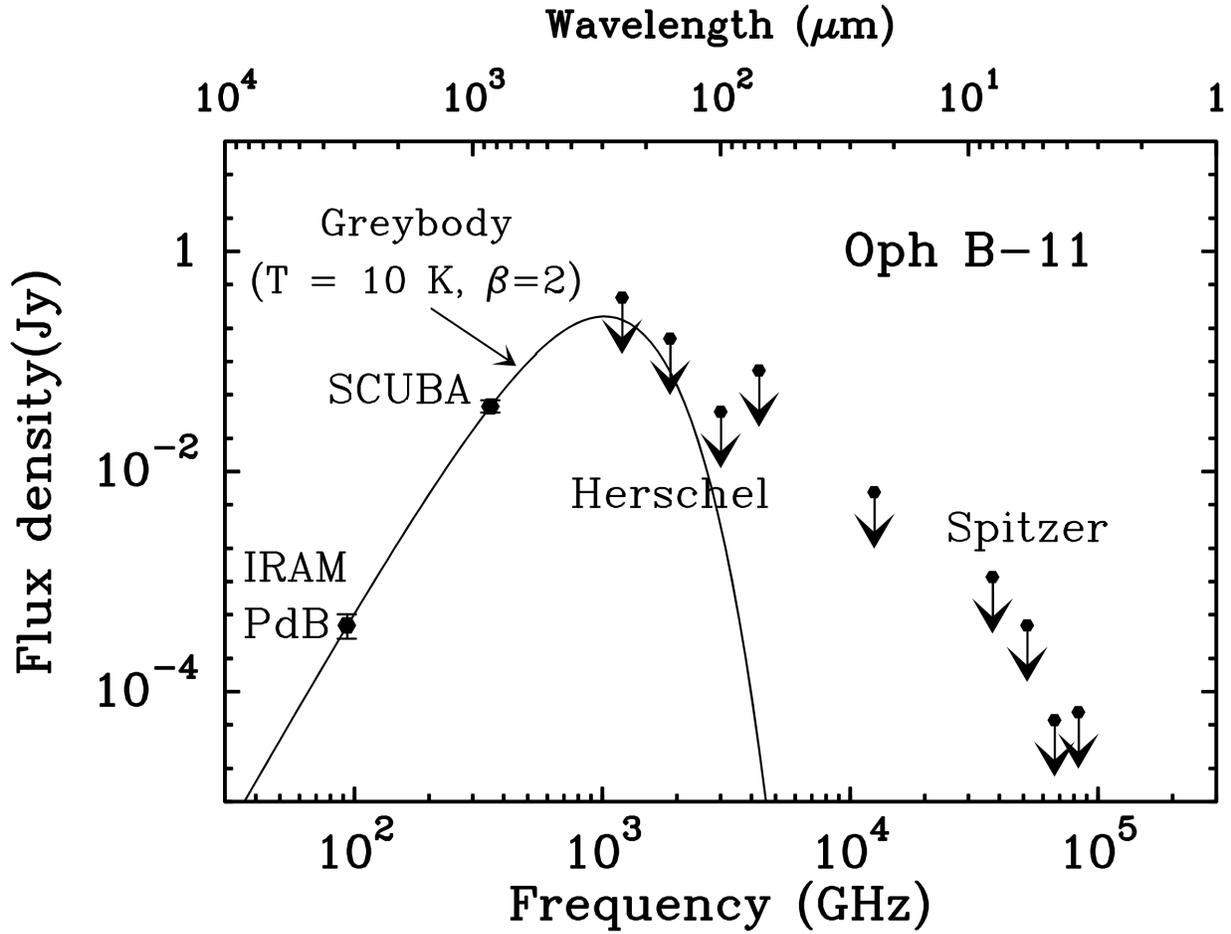

**Fig. 4.** Spectral energy distribution of the Oph B-11 core including the present IRAM PdBI 3.2mm detection, the original SCUBA 850 μm detection (*21*), and (3σ) upper limits at other wavelengths based on far-infrared data from the *Herschel* Gould Belt survey (*25*) and mid-/near-infrared data from *Spitzer* (*24*). Error bars on detections are +-1σ. The solid curve shows a 10 K greybody consistent with all observational constraints.

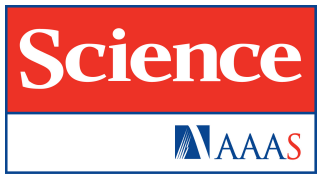

Supporting Online Material for

**Interferometric Identification of a Pre-Brown Dwarf**

Philippe André*, Derek Ward-Thompson, Jane Greaves

* To whom correspondence should be addressed. E-mail: philippe.andre@cea.fr

**This PDF file includes:**

SupplementaryText
Figs. S1 to S2
References

## Supplementary Text

**Nature of the dense gas detected in $N_2H^+$**

Figure S2A shows the detection of a prominent $N_2H^+$(1-0) core-like structure at a systemic velocity $V_{LSR}$ = 4.0 ± 0.05 km/s, which is offset by only ≲ 5" from the 3.2 mm continuum peak corresponding to Oph B-11. Based on a fitting analysis in the uv plane, the nominal FWHM diameter of this $N_2H^+$ structure is 13.2"+-0.6" (or ~ 1800 AU). The offset from the continuum peak is larger than the ≲ 1" positional uncertainty, and we interpret this offset as a result of depletion of the $N_2H^+$ molecule onto dust grains at the very high density (>>$10^6$ $H_2$ $cm^{-3}$) of Oph B-11 (*33-34*). The $N_2H^+$ observations directly trace dense gas in the immediate (< 1000 AU) vicinity of Oph B-11 and demonstrate that the nonthermal velocity dispersion of this gas is only $\sigma_{NT}$ = 0.05 ± 0.02 km/s based on the interferometer data, or at most $\sigma_{NT}$ = 0.13 ± 0.01 km/s based on the combination of the interferometer data with IRAM 30m telescope data (*35*) (cf. Fig. S2B). Moreover, the nonthermal velocity dispersion measured on the line of sight to Oph B-11 in $DCO^+$(2-1) observations taken with the IRAM 30m telescope (*35*) (cf. Fig. S1) is only $\sigma_{NT}$ = 0.16 ± 0.01 km/s. All three values are significantly smaller than the isothermal sound speed $c_s$ ≲ 0.2 km/s for a gas temperature T ≲ 10 K. The total (thermal + nonthermal) one-dimensional velocity dispersion $\sigma_{tot}$ of the ambient dense gas is thus nearly thermal. Albeit somewhat weaker than the spectrum observed at the $N_2H^+$ peak, the beam-averaged $N_2H^+$(1-0) spectrum observed toward Oph B-11, i.e., ≲ 5" from the $N_2H^+$ peak, also shows very narrow hyperfine line components (see Fig. 2), indicating that the velocity dispersion of the gas *within* the Oph B-11 core itself is also thermally dominated.

Due to likely depletion effects, the abundance of $N_2H^+$ is uncertain, making it difficult to estimate a mass from the $N_2H^+$ map. Assuming LTE and a typical excitation temperature of 10 K, the total mass of the $N_2H^+$ structure is in the range ~ 0.01-0.03 $M_\odot$ assuming an $N_2H^+$ abundance in the range between $7 \times 10^{-10}$ (*34*) and $2.5 \times 10^{-10}$ (*36*). Adopting an outer radius of 1800 AU, the virial mass of the $N_2H^+$ structure is estimated to be ~ 0.2 $M_\odot$, which is an order of magnitude larger than the mass derived from the $N_2H^+$ emission. Despite the large uncertainties, this suggests that the $N_2H^+$ structure traces unbound material surrounding Oph B-11 itself.

**Comparison with the gravo-turbulent fragmentation scenario of core formation**

Oph B-11 coincides with a local maximum in column density traced by the 3.2 mm dust continuum point source of Fig. 1 and is closely associated with a local maximum in both $N_2H^+$(1-0) emission (Fig. S2A) and $DCO^+$(2-1) emission (Fig. S1) at $V_{LSR}$ ~ 4.0 km/s. It also lies within a broad minimum in $N_2H^+$ and $DCO^+$ linewidth, where the total line-of-sight velocity dispersion of the gas is nearly sonic, i.e., $\sigma_{LOS} \simeq c_s$ ~ 0.2 km/s (see Fig. S2D). Furthermore, Oph B-11 is located only ~ 20'' (or ~ 3000 AU) away from an elongated region where two velocity components at $V_{LSR}$ ~ 3.2 km/s and $V_{LSR}$ ~ 4.0 km/s are detected in both $N_2H^+$(1-0) and $DCO^+$(2-1) [see Fig. S2C and Fig. 6b of (*35*)] and where the total line-of-sight velocity dispersion is supersonic, $\sigma_{LOS} \simeq 2.5 c_s$ (see Fig. S2D). These observed characteristics agree well with the predictions of the gravo-turbulent scenario of core formation [cf. (*37-38*)]. The 3.2 km/s velocity component corresponds to the $V_{LSR}$ of the faint DCO$^+$(3-2) feature originally detected close to Oph B-11 by (*20*) in their low-resolution (30'') single-dish search for pre-brown dwarfs in L1688. In the context of the gravo-turbulent fragmentation scenario (*15, 17*), we interpret the region of double-peaked velocity emission as an interaction zone where the ambient gas is being shocked and compressed. We suggest that the pre-brown dwarf core Oph B-11 may have formed as a result of this compression (*39*). In particular, we note that the velocity pattern seen around Oph B-11 in Fig. S2D is very reminiscent of the numerical simulations of (*37*) and (*38*). Indeed, these simulations feature velocity jumps ≲ 1 km/s and local maxima in velocity dispersion at small but finite offsets (≲ 10000 AU or ≲ 0.05 pc) from dense cores formed by turbulent shock compression [compare Fig. S2D with Fig. 1 of (*37*)]. We stress that the shocks involved here are non-dissociative, low-velocity shocks of a few km/s at most which do not destroy molecules such as DCO$^+$ or N$_2$H$^+$ (*40*).

**Comparison with ejection models of brown dwarf formation**

In the ejection scenario described by (*9-10*), protostellar embryos are ejected from unstable multiple protostellar systems at the beginning of the protostellar phase. In this picture, the ejected embryos should have LSR velocities offset from the parent cloud systemic velocity and should be observed as low-mass protostellar objects. In the hybrid model presented by (*12*), which combines disk fragmentation and subsequent ejection, the ejected objects correspond to first hydrostatic protostellar cores (*26-27*) surrounded by low-mass gaseous envelopes and the predicted ejection velocities are 0.8+-0.35 km/s. These models do not fit well the properties of Oph B-11. First, the observed LSR velocity ~ 4.0 km/s agrees to better than 0.2 km/s with the systemic velocities of other prestellar and protostellar objects in the vicinity of Oph B-11 (see Fig. S1). Second, the non detection by *Herschel*/PACS at 70 and 100 µm (Fig. 4) is a strong argument against Oph B-11 having reached the first protostellar core stage (*26-27*). Assuming an edge-on configuration, the stringent upper limits set by *Herschel* at 70 µm and 100 µm may still be marginally consistent with the fainter emission expected from a first protostellar core forming in a ~ 0.1 M$_\odot$ gaseous clump (*41*), as is the case in the hybrid scenario (*12*). However, it should be noted that this scenario predicts the presence of a diffuse power-law envelope around each ejected protostellar embryo [see Fig. 2 of (*12*)], which is ruled out in the case of Oph B-11 (see Fig. 3). We conclude that Oph B-11 is unlikely to have formed by one of the published ejection models.

**Probability of chance coincidence with an extragalactic source**

The systemic velocity of the N$_2$H$^+$(1-0) emission detected toward Oph B-11 ($V_{LSR}$ = 4.03 ± 0.04 km/s – see Fig. 2) is consistent with that expected from an object embedded in the dense clumps Oph B1 and Oph B2 of the L1688 cloud (*22, 35*). However, since the continuum source is only detected longward of 850 µm and slightly offset from the N$_2$H$^+$(1-0) peak (see Fig. S2A), the question arises as to whether it could be an extragalactic background object, seen by chance close to the $N_2H^+$(1-0) source in projection. Source counts based on SCUBA observations of the Hubble Deep Field (*42*) indicate that the number of submillimeter galaxies brighter than $S_{850\mu}$ = 35 mJy at 850 μm is about 7 deg$^{-2}$. Oph B-11 was initially detected at 850 μm through SCUBA observations of a ~ 3' diameter field (*21*), within which we estimate the presence of a maximum of ~ 10 small-scale $N_2H^+$(1-0) peaks based on single-dish $N_2H^+$ observations of L1688 (*35*) and interferometric $N_2H^+$ observations of a similar cluster-forming region (*43*). The probability of finding a SCUBA extragalactic source with $S_{850\mu}$ > 35 mJy less than 5'' away from a small-scale $N_2H^+$(1-0) peak in a field containing 10 such peaks is estimated to be only P = 7 × 10 × π × (5/3600)$^2$ ~ 4.2 ×10$^{-4}$. The hypothesis that Oph B-11 is an extragalactic source can therefore be rejected at a very high confidence level.

**Uncertainties in the mass estimate**

The mass range given for Oph B-11 in the main text ($M_{obs}$ ~ 0.02 - 0.03 $M_\odot$) already takes into account the uncertainty in the dust temperature. Apart from the latter, the main additional sources of uncertainty come from our assumed value of the dust opacity ($\kappa_{850\mu}$ ~ 0.01 cm$^2$ g$^{-1}$), which is uncertain by a factor of ~ 2, and our adopted distance to the Ophiuchus cloud [d = 140 pc from (*19*)], which is known to ~ 15%. Note that the assumptions we have made are conservative. In particular, the dust opacity model with coagulated grains of (*44*), often advocated for dense protostellar cores, has $\kappa_{850\mu}$ = 0.0185 cm$^2$ g$^{-1}$ and would thus lead to a lower mass for Oph B-11 by a factor of 1.85. Likewise, had we adopted a distance of 120 pc (*18*), we would have derived a ~ 30% lower mass. Therefore, the conclusion that the mass of Oph B-11 is substellar is robust.

**Future evolution of the core mass**

The present mass ~ 0.02-0.03 $M_\odot$ of Oph B-11 is smaller than the brown dwarf limit of 0.075 $M_\odot$, but the dense core lies in a background of relatively high density gas ($n_{H2,back}$ ≲ 5×10$^5$cm$^{-3}$ - cf. Figs. S1 and S2A), and it is possible that it will grow somewhat in mass through Bondi-Hoyle accretion of ambient material (*45*). However, the corresponding accretion rate which can be estimated as $dM_{acc}/dt \approx \pi\, \rho_{back}\, v_{rel}\, R_{acc}^2$, where $\rho_{back}$ is the background gas density, $v_{rel}$ is the relative velocity between the core and the local ambient gas, and $R_{acc} = 2GM_{core}/(v_{rel}^2 + c_s^2)$ is the Bondi-Hoyle accretion radius, cannot be larger than ~ 10$^{-7}$ $M_\odot$yr$^{-1}$ for a typical value of $v_{rel}$ ~ 0.3 km/s (*35*) given the small observed core mass. If we adopt $v_{rel}$ ~ 0.8 km/s, corresponding to the observed difference between the systemic velocity of Oph B-11 and the second velocity component detected ~ 3000 AU to the south-west of Oph B-11 (see Fig. S2 A & C), then the Bondi-Hoyle accretion rate is even smaller ~ 10$^{-8}$ $M_\odot$yr$^{-1}$. Consequently, the Bondi-Hoyle accretion timescale $t_{acc}$ = $M_{core}$/($dM_{acc}$/dt) ≳ 3×10$^5$ yr is necessarily much longer than the free-fall time $t_{ff}$ ~ 10$^4$ yr ×($<n_{H2}>$/10$^7$ cm$^{-3}$)$^{-1/2}$ of the core. While Oph B-11 may not be in free-fall collapse, its lifetime is unlikely to be longer than 3×$t_{ff}$ ~ 3×10$^4$ yr given its high mean density (*8*) >> 10$^6$cm$^{-3}$. Therefore, we estimate that the pre-brown dwarf core cannot grow by more than ~ 1/5 of its present mass before collapsing to a stellar-like object. On the other hand, some mass loss is likely to occur at the protostellar stage due to the effect of the bipolar outflow, known to be present in young brown dwarfs (*46*). As a result, only ~ 25%-75% of the core mass may end up in the final star-like object (*47*). Altogether, Oph B-11 is thus likely to form a brown dwarf with a mass ~ 0.01-0.04 $M_\odot$, within a factor of ≲ 2 of the presently observed core mass.

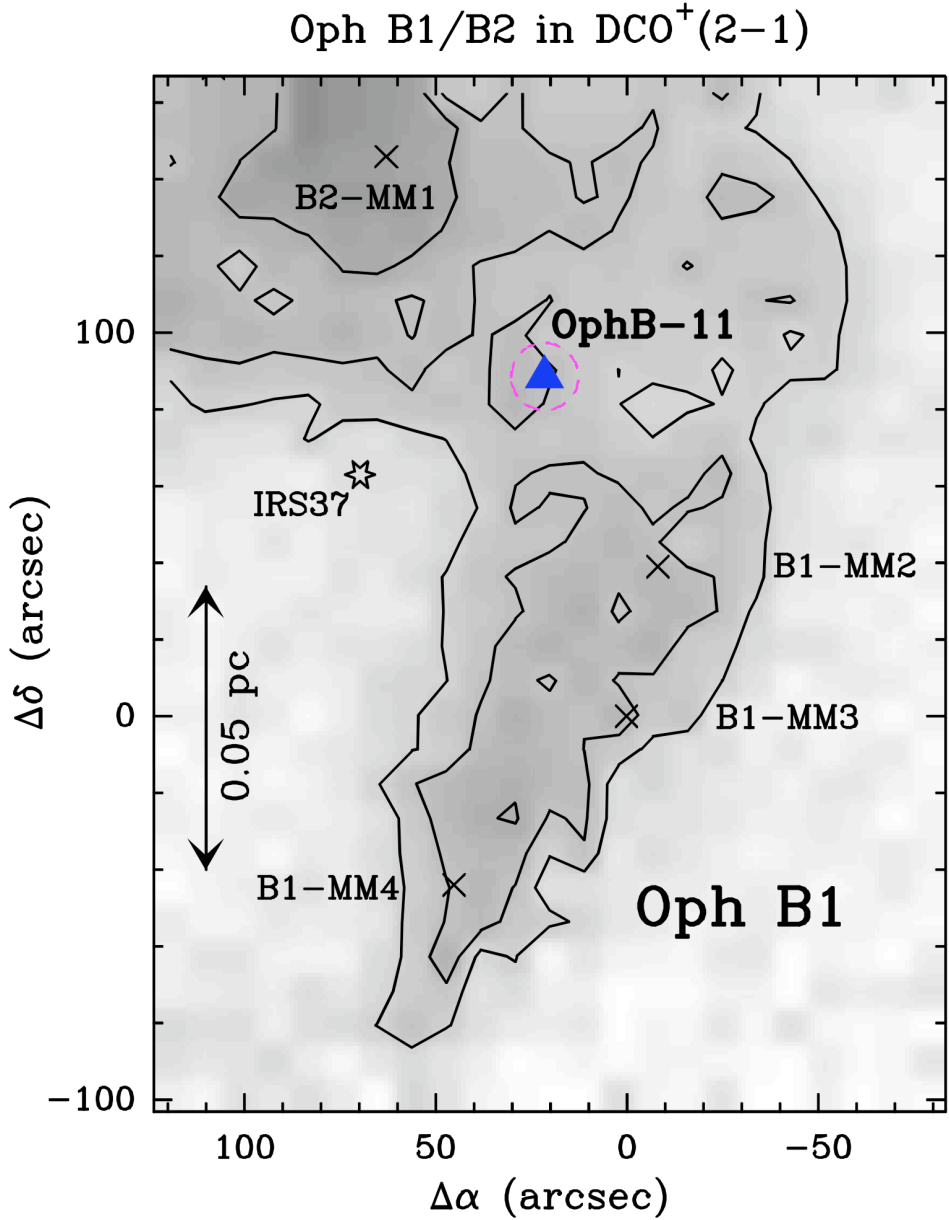

**Fig. S1.** DCO$^+$(2-1) channel map corresponding to a 0.16 km/s-wide velocity centered at $V_{LSR}$ = 4.06 km/s, showing the location of the pre-brown dwarf core Oph B-11 with respect to the dense clump Oph B1 in the L1688 cloud. The data (*35*) were taken in the on-the-fly mapping mode with the IRAM 30m telescope at an angular resolution of 17.5'' (half power beam width - HPBW). The filled triangle shows the PdBI 3.2 mm continuum position of Oph B-11 (cf. Fig. 1). The dashed circle marks the HPBW size (17.5'') of the IRAM 30m telescope in DCO$^+$(2-1). Note that Oph B-11 is closely associated with a local maximum in DCO$^+$(2-1) emission at this velocity ($V_{LSR}$ = 4.06 km/s). The contours are 1.9 K, 2.9 K, and 3.9 K. The 1σ noise level is 0.25 K. The background column density at the location of Oph B-11 is $N_{H2,back}$ ~ $3\times10^{22}$ cm$^{-2}$, according to both 2MASS near-infrared extinction data and *Herschel* observations (*25*). The DCO$^+$ map suggests that the density of the ambient cloud material around Oph B-11 is of order the critical

density of the DCO$^+$(2-1) transition, i.e., $5 \times 10^5$ cm$^{-3}$, and that the scale height of this dense gas is < 0.05 pc (or < 10000 AU). The star symbol marks the position of the protostar IRS 37 (*48*), while crosses mark the positions of four low-mass starless cores (B2-MM1 and B1-MM2 to MM4) detected in the 1.2 mm dust continuum (*30*). The systemic velocities of these five objects, as derived from N$_2$H$^+$(1-0) and DCO$^+$(2-1) observations (*35*), are all consistent with the 4.0 km/s LSR velocity of Oph B-11 to better than 0.2 km/s (better than 0.1 km/s in the case of IRS 37).

**A**

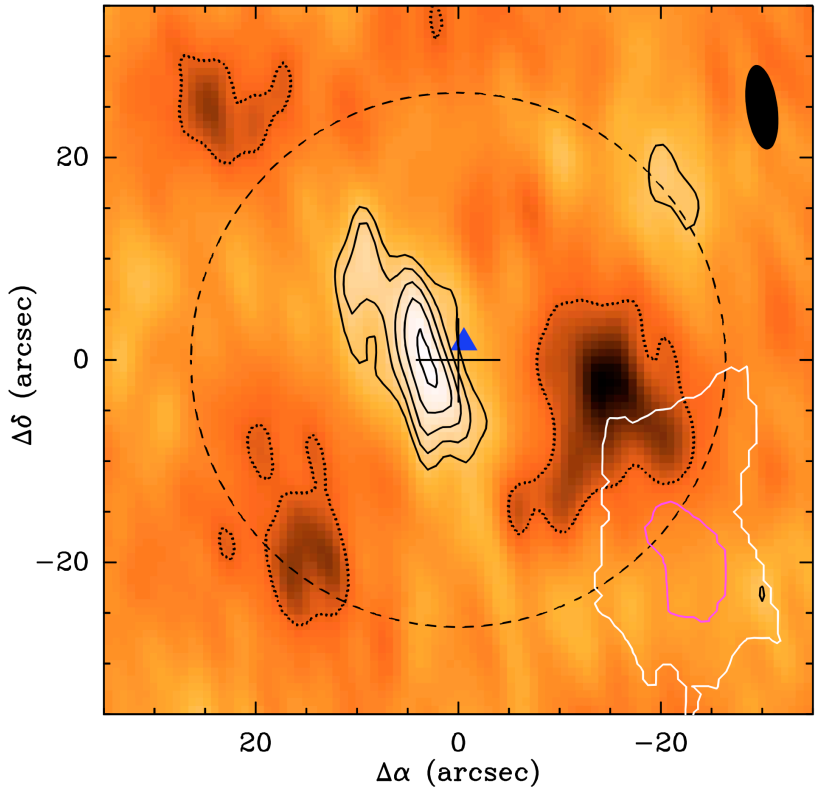

**B**

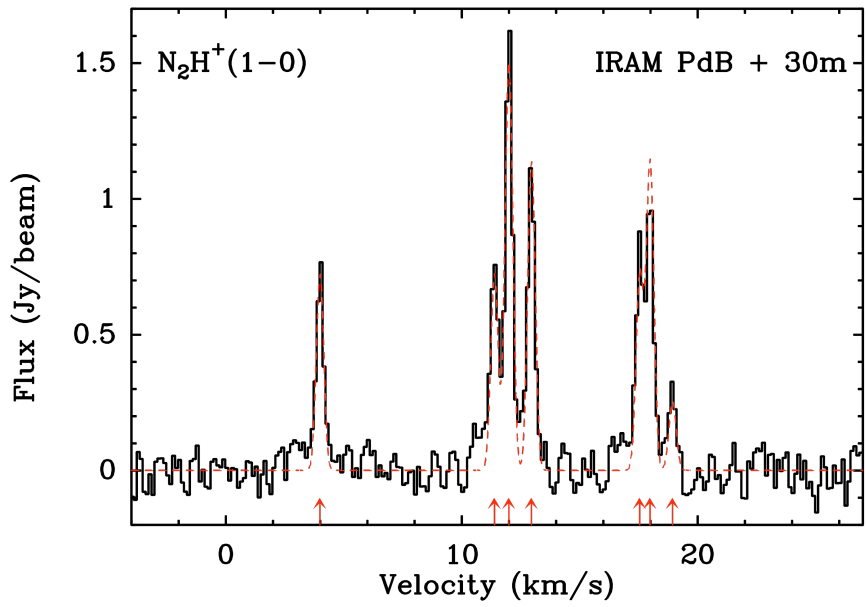

C

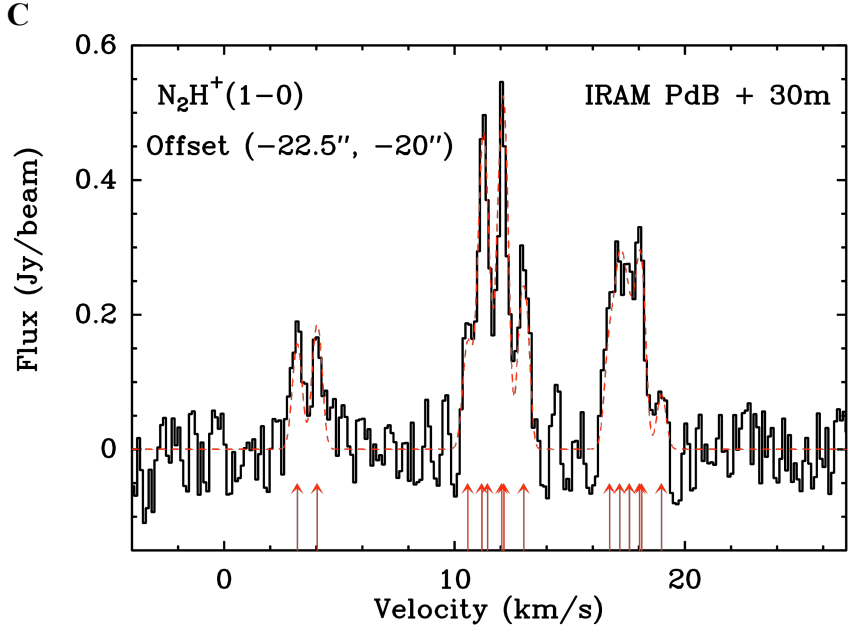

D

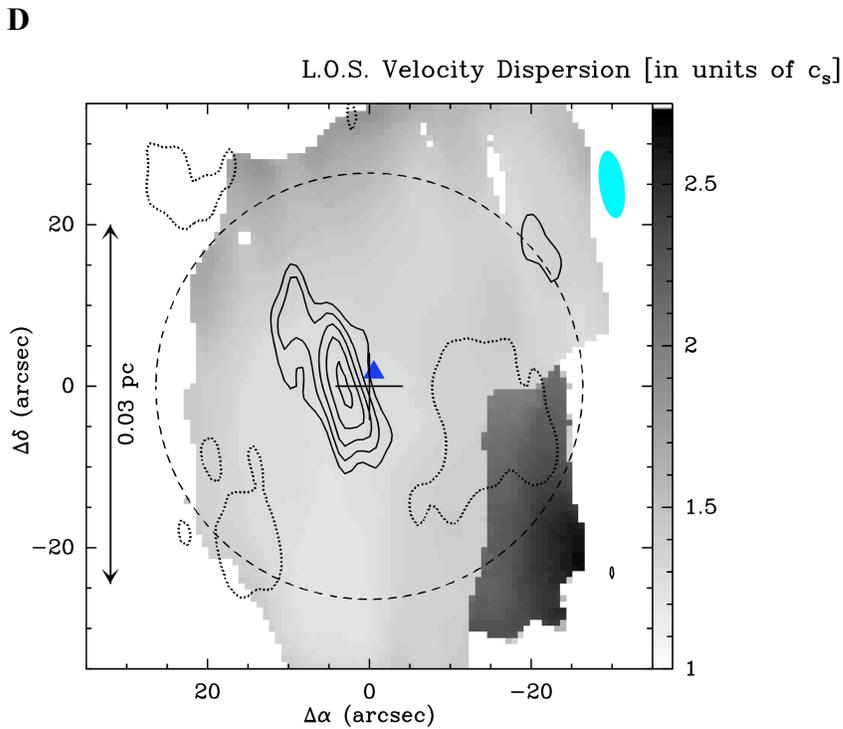

**Fig. S2.** **(A)** $N_2H^+$(101-012) image obtained toward the candidate pre-brown dwarf Oph B-11 with the CD-array of the IRAM PdBI. The $N_2H^+$(101-012) intensity has been integrated from $V_{LSR}$ = 3.85 km/s to $V_{LSR}$ = 4.2 km/s, over a velocity range centered on the systemic velocity of the $N_2H^+$ source detected with PdBI (see Fig. 2). The rms noise is $\sigma \approx 2.1$ mJy.beam$^{-1}$.km.s$^{-1}$; the contours are -3$\sigma$ (dotted), 4$\sigma$, 5$\sigma$, 6$\sigma$, 7$\sigma$, and 8$\sigma$. The SCUBA position (*21*) of Oph B-11 is marked by a cross whose size reflects the +-3'' pointing uncertainty of JCMT. The $N_2H^+$(101-012) peak is at $\Delta\alpha$ = +3.0'', $\Delta\delta$ = -0.75''

with respect to the SCUBA position. The 8.4'' × 3.1'' (HPBW) synthesized beam of the interferometer observations is shown at the upper right; the primary beam is shown as a dashed circle (diameter ~ 54''). The filled triangle shows the position of the 3.2 mm continuum peak detected with PdBI (see Fig. 1). The purple contour marks the area where a second velocity component at $V_{LSR} = 3.17 \pm 0.05$ km/s is detected above the $10\sigma$ level with PdBI, and the white contour represents the area where the same velocity component is detected above the $10\sigma$ level based on combined PdBI and IRAM 30m data.

**(B)** $N_2H^+$(1-0) spectrum resulting from combined IRAM PdBI and 30m telescope observations (*35*) at the $N_2H^+$(101-012) peak position in the upper panel. The single-dish 30m data were added to the interferometer data to provide short uv spacing information and compensate for the incomplete uv coverage of the interferometer responsible for the negative levels in the image shown in the upper panel. A Gaussian fit to the $N_2H^+$(1-0) multiplet yields a narrow FWHM linewidth $\Delta v_{obs} = 0.36 \pm 0.01$ km/s and nearly the same systemic velocity $V_{LSR} = 3.98 \pm 0.05$ km/s as for the beam-averaged $N_2H^+$ spectrum observed with PdBI toward Oph B-11 (Fig. 2). The expected positions of the seven components of the $N_2H^+$(1-0) line multiplet are marked by red arrows. Given the velocity resolution $\Delta v_{res} = 0.125$ km/s, the deconvolved FWHM linewidth is $\Delta v_{int} = 0.34 \pm 0.02$ km/s.

**(C)** $N_2H^+$(1-0) spectrum resulting from combined IRAM PdBI and 30m data (*35*) at offset $\Delta\alpha = -22.5'', \Delta\delta = -20''$ with respect to the SCUBA position of Oph B-11. Note the detection of two velocity components at this position, at $V_{LSR} = 4.03 \pm 0.05$ km/s and $V_{LSR} = 3.17 \pm 0.05$ km/s according to a two-component Gaussian fit to the $N_2H^+$(1-0) multiplet. The FWHM linewidths of these two velocity components are $\Delta v_{obs} = 0.45 \pm 0.01$ km/s and $\Delta v_{obs} = 0.49 \pm 0.01$ km/s, respectively. The expected positions of the seven components of the $N_2H^+$(1-0) line multiplet for each of the two velocity components are marked by red arrows.

**(D)** Greyscale image of the total (thermal + nonthermal) line-of-sight velocity dispersion, $\sigma_{LOS} = (\sigma^2_{NT} + c_s^2)^{1/2}$ (in units of the sound speed $c_s \simeq 0.2$ km/s), where the nonthermal component $\sigma_{NT}$ was derived from Gaussian fits to the $N_2H^+$(1-0) multiplet in our combined IRAM PdBI and 30m data. The solid contours are the same as in Fig. S2A and show the $N_2H^+$(101-012) integrated intensity emission from the Oph B-11 core as imaged by the interferometer. Note the small, nearly sonic line-of-sight velocity dispersion measured toward Oph B-11 and the large increase in velocity dispersion ~ 20'' south-west of Oph B-11 due to the presence of two velocity components there (cf. Fig. S2C). This is qualitatively very similar to the numerical simulations presented by (*37*) (compare with their Fig. 1).

**References (with titles) and Notes:**

48. B. A.Wilking, C. J. Lada, E. T. Young, IRAS observations of the Rho Ophiuchi infrared cluster - Spectral energy distributions and luminosity function. *Astrophys. J.* **340**, 823 (1989).